\documentclass[preprint,12pt]{elsarticle}

\usepackage{graphicx,url} 
\usepackage{epsfig,color} 
\usepackage{amsmath,amsfonts,amssymb}
\usepackage{theorem}

\usepackage{array}
\usepackage{multirow}

\providecommand{\tabularnewline}{\\}
\usepackage{graphics}
\usepackage{algorithm}
\usepackage{algorithmic}

\newcommand{\trace}{\mbox{\rm tr}}
\newcommand{\Prob}{\mathbb P}
\newcommand{\E}{\mathbb E}
\newcommand{\eps}{\varepsilon}
\newcommand{\Real}[1]{ { {\mathbb R}^{#1} } }

\newcommand{\sys}{{\scriptscriptstyle S}}
\newcommand{\m}{{\scriptscriptstyle M}}
\newcommand{\fsys}{f^\sys}
\newcommand{\fmod}{f^\m}

\newtheorem{example}{Example}
\newtheorem{definition}{Definition}
\newtheorem{theorem}{Theorem}
\newtheorem{remark}{Remark}
\newtheorem{algo}{Algorithm}
\newtheorem{corollary}{Corollary}
\newtheorem{assumption}{Assumption}

\begin{document}

\begin{frontmatter}

\title{Performance assessment and design of abstracted models for stochastic hybrid systems through a randomized approach
\tnoteref{ack}\tnoteref{conf}}

\tnotetext[ack]{Research supported by the European Commission under the MoVeS project, FP7-ICT-2009-257005.}
\tnotetext[conf]{A preliminary version of this work entitled
 ``A simulation-based approach to the approximation of stochastic hybrid systems'' was presented at the 4th IFAC Conference on Analysis and Design of Hybrid Systems (ADHS), Eindhoven, The Netherlands, 2012. Corresponding author M. Prandini. Tel. +39-02-23993441 Fax +39-02-23993412.}


\author[Milano]{M. Prandini}\ead{prandini@elet.polimi.it}    
\author[Milano]{S. Garatti}\ead{sgaratti@elet.polimi.it}               
\author[Milano]{R. Vignali}\ead{vignali@elet.polimi.it}  

\address[Milano]{Dipartimento di Elettronica, Informazione e Bioingegneria, Politecnico di Milano, Piazza Leonardo da Vinci, 32 - 20133 Milano, Italy}

\begin{abstract}
In this paper, a simulation-based method for the analysis and design of abstracted models for a stochastic hybrid system is proposed.  The accuracy of a model is evaluated in terms of its capability to reproduce the system output for all the realizations of the stochastic input except for a set of (small) probability $\eps$ ($\eps$-abstraction). This naturally leads to chance-constrained optimization problems, which are here tackled by means of a recently developed randomized approach. The main thrust of this paper is that, by testing how close the model and system outputs are over a finite number $N$ of input realizations only, conclusions can be drawn about the model capability as an $\eps$-abstraction. The key feature of the proposed method is its high versatility since it does not require specific assumptions on the system to be approximated. The only requirement is that of being able to run multiple simulations of the system behavior for different input realizations.
\end{abstract}


\end{frontmatter}


\section{Introduction}

This paper deals with the problem of approximating a stochastic hybrid system
by means of some simpler model, \cite{JGP06,Petreczky_Vidal_2007,JP09,A09}.
Stochastic hybrid systems (SHS) are characterized by intertwined continuous, discrete, and stochastic dynamics, and are suitable for modeling complex, large scale systems. See e.g. \cite{BLVolume,CLVolume} for an overview of applications of SHS to various domains, such as telecommunication networks, air traffic management, manufacturing, biology, finance, to mention a few. The study of SHS is more challenging than for other classes of systems, and many problems still lack an effective solution (see, for example, the motivational paper  \cite{LP2010}). In particular, this is the case of analysis and design of simple models approximating a SHS.\\
In this paper, we focus on system approximation for verification purposes. Verification of properties related to the SHS evolution, like, e.g., safety and reach/avoid properties, is typically addressed through numerical methods involving state-space gridding, \cite{AAPLS07HSCC,AKLP2010,Summers2010}, and, as such, is affected by an exponential growth of the computational effort with the state-space dimension. The aim of the approximation, then, is to build a model that mimics the behavior of the original system and that can be used in place of the system to scale-up numerical methods for the verification of the property of interest. In this respect, the notion of approximate stochastic bi-simulation as introduced in \cite{JP09,JGP06} is well-suited to quantify the model performance. \\
According to this notion, the behavior of system $S$ is characterized in terms of some output signal $y^\sys$ of interest, while model $M$ is fed with the same stochastic elements affecting the dynamics of $S$ (stochastic input and initial state) and generates a signal $y^\m$ that takes values in the same domain of $y^\sys$. The quality of $M$ as an approximate abstraction of $S$ is quantified through the maximal distance between the system and the model outputs over all possible input realizations and initial conditions except for a set of them of probability $\eps$. \\
The evaluation of the maximal distance, however, is a difficult task, computationally demanding in general. The approach proposed in \cite{JP09} is based on the quite general notion of stochastic bi-simulation function, but is able to provide a computational procedure for specific classes of SHS only. Moreover, it results in overconservative bounds as shown in the present paper.
\\
The key idea developed in this paper is to assess the quality of the approximation by resorting to randomized techniques,
which are gaining increasing popularity in the systems and control community. See \cite{CDT2011,TCD2013} for comprehensive references on the state of the art of randomized methods in the systems and control field. \\
The randomized approach proposed in this paper is in the line of the so-called scenario approach, \cite{CC05,CC06,CGP09}. It basically prescribes to compute the maximal distance between the system and the model outputs over a finite number $N$ of realizations of the initial state and of the stochastic input only. The finiteness of the considered realizations makes the problem computationally affordable. In addition, this sample-based approach is supported by a solid theory: it can be proved that if $N$ is suitably chosen depending on the desired $\eps$, then the computed distance bounds with high confidence also the distances between the outputs of $S$ and $M$ associated with all the other unseen realizations of the initial state and of the stochastic input except for a set of probability $\eps$. This idea was first mentioned in \cite{CGP09} as one of the possible applications to systems and control design of the so-called scenario approach for solving chance-constrained optimization problems. Papers \cite{AP2011,GP2012} put forward this idea, which is further elaborated here, leading to a significant improvement in terms of problem formulation, theoretical and algorithmic developments, and comparative analysis with the existing approaches.
\\
Differently from the approach based on stochastic bi-simulation functions in \cite{JP09}, the computational method here provided returns nonconservative results and is of general applicability. Indeed, the only assumption on $S$ is that one should be able to run multiple executions of $S$ and to determine the corresponding output realizations. If feasible, one could even run experiments on the real system without the need of determining a mathematical description and building a simulator for it. \\
Interestingly, the proposed framework is amenable not only for the assessment of the approximation quality of a given model, but also for model design, i.e., for selecting the best model in some given parameterized model class.
Indeed, performance assessment and model design are formulated together in the paper, being the former a special case of the latter. The problem of selecting the model class, instead, is not addressed. \\
The proposed approach should be combined with computational verification techniques to allow for the analysis of probabilistic safety and reachability properties of large scale stochastic systems. Admittedly, being based on simulation and randomization, our approach is confined to properties that depend on the system behavior over a \emph{finite horizon} and is guaranteed with a certain (arbitrarily high though) \emph{confidence}, while the method in \cite{JP09} has not such limitations.  \\
\\
\textbf{Paper structure:} We start by formulating the problem of approximating a stochastic system $S$ in Section \ref{sec:problem}, where we precisely state the issue of assessing the performance of a given abstracted model $M$ for $S$ and that of designing the abstracted model.
In Section \ref{sec:scenario}, we develop our randomized approach for both model design and performance assessment. Special focus is given in Section \ref{sec:assessment} to the performance assessment problem, which can be efficiently tackled via the scenario approach. In Section \ref{sec:design} results based on VC theory are also reported pointing out their possible conservativeness. Section \ref{sec:example} presents a numerical example where the proposed approach is compared with that in \cite{JP09}. Some final conclusions are drawn in Section \ref{sec:conclusions}. \\
\\
\textbf{Notation:} Throughout the paper, we use small letters like $s$ to denote a signal defined over the look-ahead time horizon $[0,T]$, and $s_t$ to denote the value taken by $s$ at time $t\in [0,T]$. For each $t \in [0,T]$, $s_t$ takes value in the space $\mathcal{S}$. $\mathcal{S}$ may be e.g. $\mathbb{R}^n$ or, when we are dealing with hybrid systems and $s_t$ has both a continuous and a discrete component, $\mathbb{R}^n \times \{1,2,\ldots,q\}$.
$\mathcal{S}^{[0,T]}$ denotes the set of all signals defined over
the time interval $[0,T]$ and taking values in $\mathcal{S}$ at each time instant $t\in [0,T]$.

\section{Problem formulation}\label{sec:problem}

System $S$ is described as an operator that maps the initial
state $x_0 \in \mathcal X$ and the input signal $w \in \mathcal{W}^{[0,T]}$ into the signal $y^\sys$ of interest:
\begin{align*}
&y^\sys = \fsys(x_0,w).
\end{align*}
$x_0$ and $w$ are assumed to be stochastic with
known probability measure $\Prob$. Signal $y^\sys$ takes values
in $\mathcal{Y}^{[0,T]}$. \\
Model $M$ is defined as follows
\begin{align*}
&y^\m = \fmod(x_{0},w),
\end{align*}
where $y^\m \in \mathcal{Y}^{[0,T]}$, i.e., $y^\m$ takes values in the same set of $y^\sys$. \\
Note that $S$ and $M$ are driven by the same inputs, and the aim of $M$ is that of approximating the system by producing an output $y^\m$ which is close to $y^\sys$. The fact that the map $\fmod$ depends on the initial condition $x_0$ of $S$ does not mean that the state space of $M$ has the same size as that of $S$, but that $\fmod$ incorporates the mapping from the initialization of the state of $S$ to the initialization of the (possibly lower-dimensional) state of $M$. \\
\\
To be more concrete, we here introduce the class of Jump Linear Stochastic Systems (JLSS) and present some abstracted models that can be used to approximate JLSS.

\begin{example}[JLSS]\label{example:JLSS}
Let $B$ be a Brownia motion and consider a stochastic system $S$ with state $x^\sys_t \in
\mathbb{R}^n$ that evolves within $[0,T]$ according to the
following Stochastic Differential Equation (SDE)
\begin{align}\label{eq:jump-diffusion}
&dx^\sys_t =  A x^\sys_t dt + F x^\sys_t dB_t
\end{align}
in-between the jump times $0 < \tau_1< \dots < \tau_i <
\dots\le T$ of a Poisson process $P$ with rate $\nu>0$.  At each jump time $\tau_i$, the state is reset according to
\begin{align}\label{eq:reset}
x_{\tau_i} =
(I+ R) \lim_{s\to \tau_i^-}x_{s},
\end{align}
where $I$ is the identity matrix and $R$ is a reset matrix. If $R=0$, then, no jump occurs in the state, which evolves continuously. \\
The Brownian motion $B$ is assumed to be independent of the Poisson process $P$, while both $B$ and $P$ are  assumed to be independent of the initial state $x_0 \in \mathcal X :=\mathbb{R}^n $. In this context, the stochastic input $w$ is given by the pair $(B,P)$, which takes values in $\mathcal W =\mathbb R
\times \mathbb{Z}_+$ at each time $t \in [0,T]$. The output of interest $y^\sys_t$ takes values in $\mathcal Y =\mathbb{R}^p$ and is given by
\begin{align}\label{eq:output}
y^\sys_t = C x^\sys_t.
\end{align}
This system is known as Jump Linear Stochastic System (JLSS)
since its evolution between jump times is characterized by a
SDE with drift and diffusion terms that are linear in $x^\sys_t$,
and the state resets at the jump times are linear in $x^\sys_t$ as
well. A JLSS can be seen as a SHS with a single operating mode
characterized by a SDE. When a (auto)transition occurs, the
continuous state is subject to some deterministic reset and the
continuous dynamics keeps unchanged after the transition. \\
We here present some reduced models that can be used to approximate a JLSS. \\
All models are JLSS and are characterized by a jump diffusion process $x^\m_t\in \mathbb{R}^{\tilde n}$ that satisfies the SDE
\begin{align}\label{eq:jump-diffusion-bis}
&dx^\m_t =  \tilde A x^\m_t dt + \tilde F x^\m_t dB_t,
\end{align}
and is reset according to
\begin{align}\label{eq:reset-bis}
x^\m_{\tau_i} = (I+\tilde R) \lim_{s\to \tau_i^-} x^\m_{s}
\end{align}
at the jump times $\tau_i>0$ of the Poisson process $P$.  The model output $y^\m_t\in \mathcal Y$ is given by
\begin{align}\label{eq:output-bis}
y^\m_t = \tilde C x^\m_t.
\end{align}
$\tilde A$, $\tilde F$, $\tilde R$, and $\tilde C$ are suitably defined matrices, whereas the initial condition $x^\m_0$ is a linear function of $x_0$: $x^\m_0 = L x_0$. \\
A first reduced model can be obtained by taking only a subset of the state variables $x^\sys_t$ ($\tilde n<n$) and setting $\tilde A$, $\tilde F$, $\tilde R$, and $\tilde C$ equal to suitable sub-matrices of $A$, $F$, $R$, and $C$. \\
Other reduced models can be obtained by maintaining all the state variables $x^\sys_t$ ($\tilde n=n$) and suppressing the contribution either of the Brownian motion $B$ setting $\tilde F=0$ in \eqref{eq:jump-diffusion-bis} or of the Poisson process $P$ setting $\tilde R=0$ in \eqref{eq:reset-bis}. In the first case we have
$$
\begin{cases}
dx^\m_t =  A x^\m_t dt, \quad x^\m_t \in \mathbb{R}^n \\
x^\m_0 = x_0 \\
\mbox{reset at jump time } \tau_i: \quad x^\m_{\tau_i} = (I+R) \lim_{s\to \tau_i^-} x^\m_{s}  \\
y^\m_t = C x^\m_t,
\end{cases}
$$
while in the second case
$$
\begin{cases}
dx^\m_t =  A x^\m_t dt + F x^\m_t dB_t, \quad x^\m_t \in \mathbb{R}^n \\
x^\m_0 = x_0 \\
y^\m_t = C x^\m_t.
\end{cases}
$$
\hfill \qed
\end{example}
The quality of $M$ as an approximation of $S$ is evaluated by
looking at the similarity of the output signals $y^\m$ and $y^\sys$. To this purpose, we introduce a quasi-metric
\begin{align*}
D:\mathcal{Y}^{[0,T]}\times \mathcal{Y}^{[0,T]}\to \mathbb{R}_+
\end{align*}
to assess how close signal $y^\m$ is to $y^\sys$. For example,
letting $d$ be any metric defined over $\mathcal Y$, the metric
\begin{align*}
D(y^\sys,y^\m) = \sup_{t \in [0,T]} d(y^\sys_t,y^\m_t),
\end{align*}
can be used whenever we are interested in having $y^\sys$ and $y^\m$ close to each other at each time instant. If, otherwise, we are
interested in the distance between trajectories only, the
directional Hausdorff metric can be used
\begin{align*}
& D(y^\sys,y^\m) = \sup_{t \in [0,T]} \inf_{\tau \in [0,T]} d(y^\sys_t,y^\m_\tau).
\end{align*}
As for the metric $d$, it highly depends on the space
$\mathcal{Y}$ and on the problem itself. For example, if
$\mathcal{Y} = \mathbb{R}^p$, then it is customary to use the
Euclidean metric $d(y^\sys_t,y^\m_t) = \| y^\sys_t - y^\m_t
\|$. If, instead, $\mathcal{Y} = \mathbb{R}^p \times
\{1,2,\ldots,q\}$ so that $y_t \in \mathcal{Y}$ has both a
continuous and a discrete component, say $y_t = (y^c_t,y^d_t)$,
then the metric
\begin{align*}
d(y^\sys_t,y^\m_t) =
\begin{cases}
+\infty & \mbox{if } y^{\sys,d}_t \neq y^{\m,d}_t \\
\| y^{\sys,c}_t - y^{\m,c}_t \| & \mbox{otherwise}.
\end{cases}
\end{align*}
can be used. The meaning of this metric is that we want first to check
whether $S$ and $M$ are in the same operation mode, and then,
if so, how close the continuous components of the $y$ variables
are. \\
When evaluating the quality of $M$ as an approximation of $S$,
we can require either that $y^\m$ is close to $y^\sys$ for
every and each realization of $x_0$ and $w$ or, alternatively,
that $y^\m$ is close to $y^\sys$  for all realizations of $x_0$
and $w$ except a set of them of pre-specified probability $\eps \in
(0,1)$. This latter approach is adopted in \cite{JP09} and
presents the advantage that if there exist some ``bad'' but
quite unlikely realizations that would over-penalize the
performance of $M$ as an approximation of $S$, then, they can
be discarded. Accordingly, we define the notion of $\eps$-abstraction of $S$ as follows.

\begin{definition}\label{def:eps_abstraction_h}
Model $M$ is said to be an $\eps$-abstraction of $S$ with
accuracy function $h:\mathcal X \to \mathbb R_+$ if
\begin{align} \label{eq:eps_abstraction_h}
\Prob \left\{ D \left( y^\sys,y^\m \right)^2 \leq h(x_0) \right\} \geq 1 - \eps.
\end{align}
\hfill \qed
\end{definition}
Note that, according to Definition \ref{def:eps_abstraction_h},
$D(y^\sys,y^\m)^2$  is upper bounded by some positive function $h(x_0)$ of the initial condition $x_0$. This is so because in many situations, for fixed $w$, different initializations correspond to different similarity levels of $y^\m$ and $y^\sys$ (in, e.g., linear stochastic systems, the larger $x_0$, the worse the similarity between $y^\m$ and $y^\sys$ in general), and using a uniform bound would be too conservative.  \\
In \eqref{eq:eps_abstraction_h} the approximation quality of a model is measured through $h(x_0)$ over a set of realizations of probability $1-\eps$. Evidently, the bigger $\eps$, the more $h(x_0)$ can be pushed towards small values, because $h(x_0)$ is required to be an upper bound on $D(y^\sys,y^\m)^2$ over a smaller fraction of realizations of $x_0$ and $w$. However, the approximation quality assessment in \eqref{eq:eps_abstraction_h} becomes meaningless if $\eps$ is too close to 1, and the probability $\eps$ has to be chosen so as not to penalize accuracy, while leading to sensible statements on the properties of $S$ through the analysis of $M$. This is made more explicit in Remark \ref{rmk:avoidaince_unsafe}, showing how the notion of $\eps$-abstraction can be used in system verification.

\begin{remark} \label{rmk:avoidaince_unsafe}
Suppose that $\mathcal Y=\mathbb{R}^p$ and the aim is to compute the probability that $y^\sys$ enters an unsafe set $U$, but, due to the complexity of $S$, this task is not computationally affordable using e.g. state space gridding methods on $S$, \cite{AAPLS07HSCC,AKLP2010,Summers2010}. Suppose that a model $M$ that is an  $\eps$-abstraction of $S$ with accuracy function $h(x_0)$ is available. For each initialization $x_0$, by enlarging $U$ by a width equal to $\sqrt{h(x_0)}$, a new set $\overline U(x_0)$ is obtained such that, whenever $D \left( y^\sys,y^\m \right)^2 \leq h(x_0)$, if $y^\sys$ enters $U$, then $y^\m$ enters $\overline U(x_0)$. If the abstraction $M$ is simple enough, then one can actually compute the probability that $y^\m$ enters $\overline U(x_0)$, and the probability that  $y^\sys$ enters $U$ can be upper bounded as follows:
\begin{eqnarray}
\lefteqn{ \Prob \left\{ \exists t : \; y^\sys_t \in U \right\} } \nonumber \\
& \leq & \Prob \left\{ \exists t : \; y^\sys_t \in U  | D \left( y^\sys,y^\m \right)^2 \leq h(x_0) \right\} \cdot \Prob \left\{ D \left( y^\sys,y^\m \right)^2 \leq h(x_0) \right\} \nonumber \\
& & + \Prob \left\{ D \left( y^\sys,y^\m \right)^2 > h(x_0) \right\} \nonumber \\
& \le & \Prob \left\{ \exists t : \; y^\m_t \in \overline U(x_0) | D \left( y^\sys,y^\m \right)^2 \leq h(x_0) \right\} \cdot \Prob \left\{ D \left( y^\sys,y^\m \right)^2 \leq h(x_0) \right\} \nonumber \\
& & + \Prob \left\{ D \left( y^\sys,y^\m \right)^2 > h(x_0) \right\} \nonumber \\
& \leq & \Prob \left\{ \exists t : \; y^\m_t \in \overline U(x_0) \right\} + \eps. \label{eq:bound_pr}
\end{eqnarray}
Note that considering an enlarged set $\overline U(x_0)$ whose width depends on the initialization $x_0$ may prevent the bounding in \eqref{eq:bound_pr} to be overconservative.   \qed
\end{remark}
Given Definition \ref{def:eps_abstraction_h} of $\eps$-abstraction, we next address the problems of assessing the accuracy of a given model $M$ as an $\eps$-abstraction and designing an optimal $\eps$-abstraction. In the case of the assessment of the abstraction performance, we suppose that both the operators $\fsys$ and $\fmod$ defining $S$ and $M$ are given and the objective is to assess the accuracy of $M$ as an $\eps$-abstraction of $S$. In the design of an optimal abstraction, the operator $\fmod$ defining $M$ is no more given and our goal is to choose $\fmod$ in some given class so that $M$ is an $\eps$-abstraction of $S$ with the smallest possible accuracy. Both assessment and design involve determining an accuracy function $h(x_0)$ so that condition \eqref{eq:eps_abstraction_h} is satisfied. Clearly, the solution of this problem is not unique, and we are interested in determining the ``smallest possible'' $h(x_0)$ so as to assess the actual capabilities of the model without introducing conservatism. Since $x_0$ is stochastic, the expectation of $h(x_0)$ can be taken as a sensible measure of the size of $h(x_0)$\footnote{Note that this is not the only possible choice.  One may head for alternative options, such as minimizing the maximum of $h(x_0)$ over each one of the admissible initial conditions (worst-case approach) or minimizing the value of $h(x_0)$ over all initial conditions except for a set of pre-defined probability (value-at-risk approach).}.

If we let the accuracy function and the model class be respectively parameterized by $\vartheta$ and $\lambda$, then, model design can be naturally formulated as the following optimization problem:
\begin{align}
& \min_{\vartheta,\lambda} \E[h_\vartheta(x_0)] \label{eq:cc-opt} \\
& \text{\rm subject to: } \Prob \left\{ D \left( y^\sys,y^\m_\lambda \right)^2 \leq
h_\vartheta(x_0) \right\} \geq 1 - \eps, \nonumber
\end{align}
where $y^\m_\lambda$ is the output of the parametric model. Model quality assessment can be viewed as a particular case of problem \eqref{eq:cc-opt}, where the only optimization variable is $\vartheta$.
\begin{remark}
Note that if one is dealing with model quality assessment and the accuracy function $h_\vartheta(x_0)$ is assumed to be
constant, then problem \eqref{eq:cc-opt} reduces to
\begin{align*}
& \min_{h \in \mathbb R} h \\
& \text{\rm subject to: } \Prob \left\{ D \left( y^\sys,y^\m \right)^2 \leq h
\right\} \geq 1 - \eps, \nonumber
\end{align*}
which was previously considered in \cite{AP2011} and can be
seen as a particular case of our setting. \hfill \qed
\end{remark}
The optimization problem \eqref{eq:cc-opt} is called \emph{chance-constrained} problem since we have to minimize a cost function subject to a constraint which holds in probability. Unfortunately, the constraint  $\Prob \left\{ D
\left( y^\sys,y^\m_\lambda \right) \leq h_\vartheta(x_0) \right\} \geq 1 - \eps$ is in general non-convex even when, for every fixed  realization of $x_0$ and $w$, the constraint $D \left( y^\sys,y^\m_\lambda \right) \leq h_\vartheta(x_0)$ is convex with respect to the optimization variables. For this reason, chance-constrained problems are usually hard to solve and, indeed, they are NP-hard with few exceptions, \cite{Prekopa_1996,Prekopa_2003}. In the next section, suitable algorithms aiming at finding an approximate solution to \eqref{eq:cc-opt} at low computational cost are introduced. For the sake of comparison, we first review the approach proposed in \cite{JP09}, spotting out advantages and drawbacks.

\subsection{The stochastic bi-simulation function method: a brief review}

In \cite{JP09} a method is proposed for finding a $h(x_0)$ which satisfies the probabilistic constraint \eqref{eq:eps_abstraction_h}. This method is based on the introduction of a so-called stochastic bi-simulation function and it applies to systems/models which admits a state-space representation:
$$
\begin{array}{rcl|rcl}
x^\sys & = & \phi^\sys(x_0,w) & x^\m & = & \phi^\m(x_0,w) \\
y^\sys_t & = & \psi^\sys(x^\sys_t) & y^\m_t & = & \psi^\m(x^\m_t)
\end{array},
$$
with $x^\sys_0 = x_0$ and $x^\m_0 = l(x_0)$ for some function $l$. \\
A stochastic bi-simulation function of $S$ by $M$ is a function $\pi: \mathcal X^\sys \times \mathcal X^\m  \to \Real{}_{+} \cup +\infty$ such that:
\begin{itemize}
\item[1.] $\pi(x^\sys_t,x^\m_t) \ge d(\psi^\sys(x^\sys_t),\psi^\m(x^\m_t))^2$, for any value taken by $x^\sys_t$ and $x^\m_t$;
\item[2.] the stochastic process $\pi(x^\sys_t,x^\m_t)$ is a super-martingale.
\end{itemize}
The interest in stochastic bi-simulation functions lies on the fact that, once such kind of function is found, then it is easy to prove that
\begin{align}\label{eq:SSF-accuracy}
\Prob \left\{ \left( \sup_{t \geq 0} d(y^\sys_t,y^\m_t) \right)^2 \leq \frac{\pi(x_0,l(x_0))}{\eps} \right\} \geq 1 - \eps,
\end{align}
i.e. $M$ is an $\eps$-abstraction of $S$ according to the the $\sup_{t \geq 0} d(y^\sys_t,y^\m_t)$ metric with accuracy function $\frac{\pi(x_0,l(x_0))}{\eps}$. \\
Plainly, a main issue then is that of finding a stochastic bi-simulation function for given $S$ and $M$. In \cite{JP09}, this problem is tackled for two classes of systems/models, namely the JLSS described in Example \ref{example:JLSS} and the Linear Stochastic Hybrid Automata (LSHA). It is e.g. shown that when $S$ and $M$ are both JLSS, one can consider quadratic bi-simulation functions of the type:
$$
\pi(x^\sys_t,x^\m_t) =
\begin{bmatrix}
(x^\sys_t)^T & (x^\m_t)^T
\end{bmatrix}
Q
\begin{bmatrix}
x^\sys_t \\
x^\m_t
\end{bmatrix}.
$$
When $d$ is the Euclidean metric, conditions 1. and 2. then translates into the following Linear Matrix Inequalities:
\begin{eqnarray*}
Q - \mathbf{C}^T \mathbf{C} & \succcurlyeq & 0 \\
Q (\mathbf{A} + \nu \mathbf{R}) + (\mathbf{A} + \nu \mathbf{R})^T Q + \mathbf{F}^T Q \mathbf{F} + \nu\mathbf{R}^T Q \mathbf{R} & \preccurlyeq & 0,
\end{eqnarray*}
where
we recall that $\nu$ is the rate of the Poisson process, whereas $\mathbf{C}$, $\mathbf{A}$, $\mathbf{R}$, and $\mathbf{F}$ are given by
 $$
\mathbf{C} =
\begin{bmatrix}
C & - \tilde C
\end{bmatrix}, \quad
\mathbf{A} =
\begin{bmatrix}
A & 0 \\
0  &  \tilde A
\end{bmatrix}, \quad
\mathbf{R} =
\begin{bmatrix}
R & 0 \\
0  &  \tilde R
\end{bmatrix}, \quad
\mathbf{F} =
\begin{bmatrix}
F & 0 \\
0  &  \tilde F
\end{bmatrix},
$$
with matrices $C$, $A$, $R$, $F$ and $\tilde C$, $\tilde A$, $\tilde R$, $\tilde F$ describing $S$ (see equations \eqref{eq:jump-diffusion}, \eqref{eq:reset} and \eqref{eq:output}) and $M$ (see equations \eqref{eq:jump-diffusion-bis}, \eqref{eq:reset-bis} and \eqref{eq:output-bis}), respectively.
Then, setting $x_0^\m = l(x_0^\sys)=L x_0^\sys$, problem
\begin{align}\label{eq:ass-SSF}
& \min_{Q\succcurlyeq 0} \E \left[ \begin{bmatrix}
x^T_0 & x^T_0 L^T
\end{bmatrix}
Q
\begin{bmatrix}
x_0 \\
L x_0
\end{bmatrix} \right] \\
& \text{\rm subject to: } \nonumber\\
& Q - \mathbf{C}^T \mathbf{C} \succcurlyeq 0 \nonumber\\
& Q (\mathbf{A} + \nu \mathbf{R}) + (\mathbf{A} + \nu\mathbf{R})^T Q + \mathbf{F}^T Q \mathbf{F} + \nu\mathbf{R}^T Q \mathbf{R} \preccurlyeq 0 \nonumber
\end{align}
can be solved to optimize the accuracy function $h(x_0)= \frac{\pi(x_0,l(x_0))}{\eps}$ in equation \eqref{eq:SSF-accuracy}. Note that, though this was not considered in \cite{JP09}, problem \eqref{eq:ass-SSF} can be extended to address the design of $M$ by introducing further optimization variables representing some parametrization of $M$.
\\
Despite its elegance, the bi-simulation approach suffers from the following drawbacks:
\begin{itemize}
\item[-] It is difficult to work out a bi-simulation function in general, and, in this respect, the cases of JLSS and LSHA  are more exceptions than rules. To the present state of knowledge, the usability of the bi-simulation approach is limited to very few classes of systems and models.
\item[-] Being generated by a stochastic bi-simulation function is a loose sufficient condition for $h(x_0)$ to be an accuracy function, and it may happen that $\Prob \left\{ \left( \sup_{t \geq 0} d(y^\sys_t,y^\m_t) \right)^2 > \frac{\pi(x_0,l(x_0))}{\eps} \right\}$ is much smaller  than $\eps$. Hence, there are no guarantees about the optimality of the obtained $h(x_0)$ with respect to the condition \eqref{eq:eps_abstraction_h}. This may lead to a severe underestimation of the abstraction capabilities of $M$ and eventually to conservative results.
\end{itemize}

\section{A new method based on randomization} \label{sec:scenario}

In recent years, a considerable effort has been devoted to the development of the scenario approach, a randomized algorithm for the resolution of chance-constrained problems, see e.g. \cite{CC05,CC06,CG2008,CGP09,ATC2009,CG2011,GC2013,BBHPDB2013}. The scenario approach allows the user to find approximate yet guaranteed solutions at relatively low computational effort. Here, we rely on this method to tackle problem \eqref{eq:cc-opt}. \\
Algorithmically speaking, the scenario approach builds on a very intuitive and basic idea: a number,
say $N$, of realizations of $x_0$ and $w$, say $x_0^{(i)}$ and
$w^{(i)}$ for $i=1,2,\ldots,N$, are extracted according to the
underlying probability measure $\Prob$ and optimization is
performed by taking into account this finite number of instances of $x_0$ and $w$ only. More precisely, letting $\alpha$ be a user chosen parameter such that $0 \leq \alpha < \eps$, and letting $y^{\sys,(i)} = \fsys(x_0^{(i)},w^{(i)})$ and $y^{\m,(i)}_\lambda =
\fmod_\lambda(x_0^{(i)},w^{(i)})$, $i=1,2,\ldots,N$, the randomized algorithm described in the following Algorithm~\ref{algo:scenario_algo} aims
at finding a solution that violates the constraint
\begin{align*}
D \left( y^{\sys,(i)},y^{\m,(i)}_\lambda \right)^2 \leq h_\vartheta(x_0^{(i)})
\end{align*}
$\lfloor \alpha N \rfloor$ times\footnote{$\lfloor \cdot \rfloor$ denotes integer part.} out of $N$, that is, with an \emph{empirical} probability equal to $\alpha$. We choose $\alpha < \eps$ because, as it  is intuitive, it is very likely that the actual probability with which the constraint $D \left( y^\sys,y^\m_\lambda \right)^2 \leq h_\vartheta(x_0)$ is violated is larger than the empirical one, and, hence, if $\alpha$ were exceeding $\eps$, then the actual violation probability could not be guaranteed to be smaller than the desired level $\eps$ as required in \eqref{eq:cc-opt}. Ideally, one should determine  $N-\lceil \alpha N \rceil$ uncertainty instances out of $N$ that result in the smallest value of the cost. Given that achieving the best possible overall cost reduction is a hard combinatorial problem, a sub-optimal solution is adopted in Algorithm \ref{algo:scenario_algo}.

\begin{algo}[Randomized Algorithm] \label{algo:scenario_algo}
$ $
\begin{itemize}
\item[0:] \verb"EXTRACT" $N$ realizations of $x_0$ and $w$: $x_0^{(i)}$ and
$w^{(i)}$, $i=1,2,\ldots,N$;
\item[1:] \verb"SET"
\begin{eqnarray*}
\vartheta^\ast,\lambda^\ast & := & \arg \min_{\vartheta,\lambda} \E[h_\vartheta(x_0)] \\
& & \text{\rm subject to: } D \left( y^{\sys,(i)},y^{\m,(i)}_\lambda \right)^2 \leq
h_\vartheta(x_0^{(i)}), \; i \in \{ 1,2,\ldots,N \};
\end{eqnarray*}
\item[2:] \verb"SET"  $V := \emptyset$ \verb"AND" $p := 0$; \\
\hfill{\% $V$ is the set of indexes of constraints violated by $\vartheta^\ast,\lambda^\ast$}\\
\hfill{\% $p$ is the cardinality of $V$}
\item[3:] \verb"WHILE" $p< \lfloor \alpha N \rfloor$
\begin{itemize}
\item[3.1:] \verb"SET" $\{i_1,i_2,\ldots,i_m\} := \left\{ i: \; D \left( y^{\sys,(i)},y^{\m,(i)}_{\lambda^\ast} \right)^2 = h_{\vartheta^\ast}(x_0^{(i)}) \right\}$;\\
 \hfill{\% $\{i_1,i_2,\ldots,i_m\}$ are the indexes of active constraints}
\item[3.2:] \verb"FOR" $k=1,2,\ldots,m$
\begin{itemize}
\item[3.2.1:] \verb"SET"
\begin{align*}
&\widehat \vartheta,\widehat \lambda  :=  \arg \min_{\vartheta,\lambda} \E[h_\vartheta(x_0)]\\
& \text{\rm subject to: } D \left( y^{\sys,(i)},y^{\m,(i)}_\lambda \right)^2 \leq
h_\vartheta(x_0^{(i)}), \\
&\hspace*{2.3cm} i \in \{ 1,2,\ldots,N \} / (\{i_k\} \cup V);
\end{align*}
\item[3.2.2:] \verb"IF" $\E[h_{\widehat \vartheta}(x_0)] < \E[h_{\vartheta^\ast}(x_0)]$ \verb"THEN SET" $\vartheta^\ast:=\widehat \vartheta$, $\lambda^\ast:=\widehat \lambda$;
\end{itemize}
\verb"END FOR"
\item[3.3:] \verb"SET" $V := \left\{ i: \; D \left( y^{\sys,(i)},y^{\m,(i)}_{\lambda^\ast} \right)^2 > h_{\vartheta^\ast}(x_0^{(i)}) \right\}$ \verb"AND" $p := |V|$;\\
\hfill{\% $V$ is the set of indexes of constraints violated by $\vartheta^\ast,\lambda^\ast$}\\
\hfill{\% $p$ is the cardinality of $V$}
\end{itemize}
\verb"END WHILE"
\item[4:] \verb"RETURN" $\vartheta^\ast,\lambda^\ast$.
\end{itemize}
\end{algo}
In the algorithm, the constraints to be violated are progressively selected by discarding one active constraint at a time, precisely, the constraint that, when removed, gives the largest immediate cost improvement (greedy approach). The search is restricted to active constraints only, because eliminating a nonactive constraint does not improve the cost value. Though the greedy approach may not yield the best possible overall cost reduction, a fair sub-optimality is obtained, while the computational effort is kept at a reasonable level.
\begin{remark}[Alternative removal rules] \label{rmk:alternative_algo}
If the greedy approach is still too computationally expensive, variants of Algorithm~\ref{algo:scenario_algo} can be adopted to further reduce the computational effort.
For instance, one can choose one constraint at random among the active ones at each iteration of the \texttt{WHILE} cycle (random removal), or
one can remove at each step all the active constraints (block removal) until the number of constraints to be removed is lower than the number of active ones, in which case the last constraints to be removed can be chosen at random among the active ones.
Though the solution achieved through these approaches is even more sub-optimal than the greedy one, notably, the guarantee on chance-constrained feasibility given in the theorems to follow continues to hold. \qed
\end{remark}
Each optimization problem that has to be solved in the Algorithm~\ref{algo:scenario_algo} is of standard type, i.e. with a finite number of constraints. In particular, if the cost function and the constraints are convex, then, the problem can be tackled via optimization modeling languages like CVX, \cite{cvx}, and YALMIP, \cite{YALMIP}, equipped with standard solvers. An inspection of the code reveals that Algorithm~\ref{algo:scenario_algo} comes to termination as long as, each time the \texttt{FOR} cycle at line \textit{3.2} is called, one active constraint whose removal improves the cost can be found. This condition is satisfied in normal situation and is assumed here for granted.\\
\\
Although obtained based on a finite number of samples of $x_0$ and
$w$ only, the solution returned by the randomized Algorithm~\ref{algo:scenario_algo} (hereafter, called the randomized solution) comes with precise guarantees about
its feasibility for the original chance-constrained problem \eqref{eq:cc-opt}.
This is the main feature of the scenario approach, which, hence, can be reliably (as opposed to empirically) used to tackle chance-constrained problems otherwise deemed intractable. \\
The following theorem precisely states this feasibility property and can be derived quite directly from \cite[Theorem 2.1]{CG2011}
under the following assumption.

\begin{assumption}[convexity]\label{ass:convexity}
$\E[h_\vartheta(x_0)]$ is a convex function of $\vartheta$ and, for every fixed realization of $x_0$ and $w$, the constraint $D \left( y^\sys,y^\m_\lambda \right)^2 \leq h_\vartheta(x_0)$ is convex in the optimization variables $\vartheta,\lambda$.
\end{assumption}

\begin{theorem}[feasibility of the randomized solution] \label{th:scenario}
$\mbox{ }$\\
Under Assumption \ref{ass:convexity}, if $N$ is big enough so that ($r$ is the overall dimensionality of the optimization variables $\vartheta,\lambda$)
\begin{align} \label{eq:findN_implicit}
{\lfloor \alpha N \rfloor + r-1 \choose \lfloor \alpha N \rfloor}
\sum_{i=0}^{\lfloor \alpha N \rfloor + r-1} {N \choose i} \eps^i
(1-\eps)^{N-i} \leq \beta,
\end{align}
then the randomized solution $(\vartheta^\ast, \lambda^\ast)$ is such that
\begin{align*}
\Prob \left\{ D \left( y^\sys,y^\m_{\lambda^\ast} \right)^2 \leq h_{\vartheta^\ast}(x_0) \right\} \geq
1 - \eps
\end{align*}
with confidence at least $1-\beta$. \hfill \qed
\end{theorem}
The theorem basically says that the randomized solution can be made feasible for \eqref{eq:cc-opt} with high confidence. To this purpose, it is worth noticing that it is not possible to guarantee that the randomized solution is always feasible for \eqref{eq:cc-opt}, since this solution depends on the $N$ extracted samples $x_0^{(i)},\ w^{(i)}$ and it may well happen that these samples are not enough representative of the whole distribution of $x_0$ and $w$. Yet, this latter case is very unlikely for large $N$ and, indeed, Theorem \ref{th:scenario} says that if $N$ is chosen as indicated, then, the probability of such bad event is no greater than $\beta$. \\
In Theorem~\ref{th:scenario}, the sample size $N$ is implicitly given. Explicit bound on the sample size can be obtained by relying on suitable inequalities for the binomial term in \eqref{eq:findN_implicit}, see \cite{C2009,ATL2010a,ATL2010b}. In particular the following corollary can be derived by using the so-called Chernoff bound, \cite{TCD2013}, in a way similar to that adopted in \cite{C2009}.

\begin{corollary} \label{corol:scenario}
Under the assumptions of Theorem~\ref{th:scenario}, if
\begin{equation} \label{eq:findN}
N \geq \frac{(2+\alpha) \eps}{(\eps-\alpha)^2} \left[ (r-1) \ln \left( \frac{2\eps (2+\alpha)  (r-1)}{(\eps-\alpha)^2} \right) + \ln \frac{1}{\beta} \right] + \frac{r-1}{2},
\end{equation}
then the randomized solution $(\vartheta^\ast, \lambda^\ast)$ is such that
\begin{align*}
\Prob \left\{ D \left( y^\sys,y^\m_{\lambda^\ast} \right)^2 \leq h_{\vartheta^\ast}(x_0) \right\} \geq
1 - \eps
\end{align*}
with confidence at least $1-\beta$. \hfill \qed
\end{corollary}
\textbf{Proof:} see \ref{appendix:1}. \\
\\
The explicit bound in \eqref{eq:findN} reveals a very important fact, namely, that $N$ increases logarithmically with $\beta$. This means that we can enforce a very small value for $\beta$ -- like $\beta = 10^{-10}$, which guarantees the achievement of $\Prob \left\{ D \left( y^\sys,y^\m \right)^2 \leq h(x_0) \right\} \geq 1 - \eps$ beyond any reasonable doubt -- without affecting the sample size $N$ too much. \\
\\
The following remark sheds light on the role of $\alpha$ as a means to tune the level of approximation of the randomized solution.

\begin{remark}[Choice of the empirical probability of violation $\alpha$] \label{remark:alpha}
It is worth noticing that the empirical probability of violation $\alpha$ is a user chosen parameter through which the level of approximation of the randomized solution can be tuned. If one chooses $\alpha=0$, then, no constraints need to be removed and the problem reduces to finding a solution to a single optimization   problem. This is computationally attractive, but the actual violation of the obtained randomized solution is typically much smaller than the desired $\eps$  and the performance of the model significantly underestimated.
As a matter of fact, though the feasibility of the randomized solution is guaranteed for every $\alpha \in [0,\eps)$, it is intuitively clear that the closer $\alpha$ to the desired violation probability $\eps$ the better the randomized solution approximates the actual solution to the chance-constrained problem \eqref{eq:cc-opt}. At the same time, however, it holds that $N \to \infty$ as $\alpha \to \eps$, see \eqref{eq:findN}. The ultimate choice for $\alpha$ rests with the user, who can select his/her own best comprise between the accuracy required by the application at hand and computational tractability.  \qed
\end{remark}
As is clear the applicability of Theorem~\ref{th:scenario} rests on the validity of Assumption \ref{ass:convexity}. It is a fact that Assumption \ref{ass:convexity} can be easily satisfied when addressing performance assessment as discussed in the next Subsection \ref{sec:assessment}. When dealing with model design, instead, the satisfaction of Assumption \ref{ass:convexity} depends on the model parametrization and it may be harder to achieve. Subsection \ref{sec:design} hints at some possible extensions of Theorem~\ref{th:scenario} to the non-convex  case.

\subsection{Performance assessment} \label{sec:assessment}

In performance assessment, the sole optimization variable is
$\vartheta$, the parameter of $h$, being $\fmod$ given and fixed. \\
In order to apply Theorem~\ref{th:scenario}, we need to
ensure the convexity with respect to $\vartheta$ of both $\E[h_\vartheta(x_0)]$ and the constraint $D \left( y^\sys,y^\m \right)^2 \leq h_\vartheta(x_0)$.
Since the convexity $\E[h_\vartheta(x_0)]$ is achieved when
$h_\vartheta(x_0)$ is convex in $\vartheta$, while the
convexity of $D \left( y^\sys,y^\m \right)^2 \leq
h_\vartheta(x_0)$ requires that $h_\vartheta(x_0)$ is concave
in $\vartheta$, function $h_\vartheta(x_0)$ must be linearly parameterized in $\vartheta$. \\
Plainly, a possible parametrization is
\begin{align*}
h_\vartheta(x_0) = \sum_{i=1}^l \vartheta_i h_i(x_0),
\end{align*}
where $h_i(x_0)$, $i=1,2,\dots,l$, are given positive basis
functions\footnote{E.g., when $x_0 \in \mathbb R^n$, i.e., the state has no discrete components, $h_i(x_0) =\exp(- (x_0 - m_i)' V_i (x_0 - m_i))$ with $m_i$ and $V_i$
given.}, subject to the linear condition $\vartheta_i \geq 0$,
$\forall i$. We suggest, however, to use an alternative parametrization, namely, the class of positive quadratic hybrid functions of the continuous part of $x_0$, which seems to fit many situations of
interest where, for each mode $x_0^d$, the approximation
capability of model $M$ is better for a certain initial
condition $x_0^c = \bar x_0^c$ and decreases as $x_0^c$ moves
away from $\bar x_0^c$. To be precise, letting $x_0 = (x_0^c,  x_0^d)$ be
the decomposition  of $x_0$ into its continuous part $x_0^c$,
taking value in $\mathbb R^n$, and its discrete part $x_0^d$,
taking value in the finite alphabet $\{1,2,\ldots,q\}$,
$h_\vartheta(\cdot)$ is parameterized as follows
\begin{align*}
& h_\vartheta(x_0)  = \sum_{k=1}^q \left[ {x_0^c}' \Theta^A_k x_0^c + 2 {\Theta^b_k} x_0^c + \Theta^c_k \right] \mathbf{1}_{[x_0^d = k]},
\end{align*}
where $\mathbf{1}_{[\cdot]}$ is the indicator function and $\vartheta$ is the vector of the entries of $\Theta^A_k$, $\Theta^b_k$, $\Theta^c_k$, $k=1,\ldots,q$. \\
Letting
\[
\Theta_k = \begin{bmatrix}
    \Theta^A_k & {\Theta^b_k}' \\
    {\Theta^b_k} & \Theta^c_k
\end{bmatrix},
\]
then, we have that
\[
{x_0^c}' \Theta^A_k x_0^c + 2 {\Theta^b_k} x_0^c + \Theta^c_k ={\begin{bmatrix} {x_0^c}' & 1\end{bmatrix}} \Theta_k
\begin{bmatrix}x_0^c \\ 1 \end{bmatrix},
\]
and the condition of positiveness of $h_\vartheta(x_0)$ simply
translates into a positive semi-definite condition on the
matrices $\Theta_k$, that is, $\Theta_k \succeq 0$,
$k=1,2,\ldots,q$, which is a convex constraint on $\Theta_k$.
Moreover, $\E\left[h_\vartheta(x_0)\right]$ can be expanded as
follows ($\trace$ denotes trace):
\begin{align*}
& \E\left[h_\vartheta(x_0)\right] = \sum_{k=1}^q \E\left[ \trace\left( \begin{bmatrix} {x_0^c}' & 1\end{bmatrix} \Theta_k
\begin{bmatrix}x_0^c \\ 1 \end{bmatrix} \mathbf{1}_{[d_0 = k]} \right) \right]  \\
& \; = \sum_{k=1}^q \E\left[ \trace\left(  \Theta_k
\begin{bmatrix}x_0^c \\ 1 \end{bmatrix}\begin{bmatrix} {x_0^c}' & 1\end{bmatrix} \mathbf{1}_{[x_0^d = k]} \right) \right]\\
& \; = \sum_{k=1}^q \trace\left(  \Theta_k
\E\left[\begin{bmatrix}x_0^c {x_0^c}' &{x_0^c} \\ {x_0^c}'& 1 \end{bmatrix} \mathbf{1}_{[x_0^d = k]}  \right]\right)\\
& \; = \sum_{k=1}^q \trace\left(  \Theta_k
\E\left[\begin{bmatrix}x_0^c {x_0^c}' &{x_0^c} \\ {x_0^c}'& 1 \end{bmatrix} \Big|x_0^d = k   \right] \Prob(x_0^d = k ) \right),
\end{align*}
where the conditional expectation in the last equality can be
computed from the knowledge of $\Prob$. When $x_0
\in \mathbb R^n$, i.e. the state has no discrete component,
then the parametrization simplifies to $h_\vartheta(x_0) =
{\begin{bmatrix} {x_0}' & 1\end{bmatrix}} \Theta
\begin{bmatrix}x_0 \\ 1 \end{bmatrix}$, $\Theta
\succeq 0$, while $\E [ h_\vartheta(x_0) ] = \trace \left(
\Theta \E\begin{bmatrix}x_0 {x_0}' &{x_0} \\
{x_0}'& 1 \end{bmatrix} \right)$. \\
The overall implementation of the randomized algorithm for abstraction performance assessment with the parametrization of $h_\vartheta$ discussed in this section consists of the following steps:

\begin{itemize}
\item[1:] Choose $\eps \in (0,1)$, $\beta \in (0,1)$, and $\alpha \in
[0,\eps)$. Let $N$ be the smallest integer satisfying \eqref{eq:findN_implicit} (or, alternatively, let $N$ be the smallest integer satisfying \eqref{eq:findN}).

\item[2:] Extract $N$ realizations of the stochastic input $w^{(i)}$,
$i=1,2,\dots,N$, and of the initial condition $x_0^{(i)}$,
$i=1,2,\dots,N$.

\item[3:] Run the corresponding $N$ executions of $S$ and $M$ to
compute via simulation $N$ realizations of the output signals
\begin{align*}
& y^{\sys,(i)}=\fsys(x_0^{(i)},w^{(i)}), & i=1,2, \dots,N \\
& y^{\m,(i)}=\fmod(x_0^{(i)},w^{(i)}),  & i=1,2, \dots,N.
\end{align*}
Compute $D(y^{\sys,(i)},y^{\m,(i)})$, $i=1,2,\dots,N$.

\item[4:] Run Algorithm~\ref{algo:scenario_algo} with the following objective
\begin{align*}
\min_{\Theta_1 \succeq 0,\dots,\Theta_q \succeq 0} \sum_{k=1}^q \trace\left(  \Theta_k
\E\left[\begin{bmatrix}x_0^c {x_0^c}' &{x_0^c} \\ {x_0^c}'& 1 \end{bmatrix} \Big|x_0^d = k   \right] \Prob(x_0^d = k ) \right)
\end{align*}
and constraints
\begin{align*}
D(y^{\sys,(i)},y^{\m,(i)})^2 \leq \sum_{k=1}^q {\begin{bmatrix} {x_0^{c,(i)}}' & 1\end{bmatrix}} \Theta_k
\begin{bmatrix}x_0^{c,(i)} \\ 1 \end{bmatrix} \mathbf{1}_{[x_0^{d,(i)} = k]}.
\end{align*}
\end{itemize}

\begin{remark}
As for the dependence of the computational effort on the size $n$ of the continuous state component, since matrices $\Theta_k$, $k=1,\dots,q$, are symmetric and of size n+1, it follows from Corollary \ref{corol:scenario} that the number $N$ of realizations involved in the implementation of the randomized algorithm scales as $n^2\ln(n)$. \qed
\end{remark}

\subsection{Some hints for addressing the non-convex case}\label{sec:design}

Though convexity is advantageous from a computational perspective, admittedly, relying on Theorem~\ref{th:scenario} only may be limitative in our context because it is often the case that the constraint $D \left( y^\sys,y^\m_\lambda \right)^2 \leq h_\vartheta(x_0)$ is not convex, especially because of the dependence on $\lambda$. We here hints at some results that can be used in the non-convex case. Though these results are not conclusive, because of the inherent difficulty of this case, they may be useful for some problems, and, moreover, they represent a promising start for future research.

The following theorem can be derived from \cite[Theorem 7]{ATC2009} and provides guarantees about the chance-constrained feasibility of the randomized solution under a condition other than convexity. We need a preliminary definition.

\begin{definition}
For given $N$ realizations $(x_0^{(1)},w^{(1)}), \ldots,(x_0^{(N)},w^{(N)})$ of $x_0$ and $w$, let $\phi(x_0^{(1)},w^{(1)}, \ldots,x_0^{(N)},w^{(N)})$ denote the number of distinct binary vectors of the type
$$
\left[ \mathbf{1}_{D \left( y^{\sys,(1)},y^{\m,(1)}_{\lambda} \right)^2 \leq h_{\vartheta}(x_0^{(1)})} \; \mathbf{1}_{D \left( {y^\sys,(2)}, y^{\m,(2)}_{\lambda} \right)^2 \leq h_{\vartheta}(x_0^{(2)})} \; \cdots \; \mathbf{1}_{D \left( y^{\sys,(N)},y^{\m,(N)}_{\lambda} \right)^2 \leq h_{\vartheta}(x_0^{(N)})} \right]^T
$$
that are obtained while letting $\vartheta,\lambda$ vary in $\Real{r}$ ($\mathbf{1}$ denotes the indicator function). Moreover, let
$$
\pi(N) = \sup_{(x_0^{(1)},w^{(1)}), \ldots,(x_0^{(N)},w^{(N)})} \phi(x_0^{(1)},w^{(1)}, \ldots,x_0^{(N)},w^{(N)}).
$$
The VC dimension associated to Problem \eqref{eq:cc-opt} is denoted by $d_{VC}$ and is the largest integer $N$ such that the equality $\pi(N) = 2^N$ is satisfied.
\end{definition}

\begin{theorem} [feasibility of the randomized solution -- non-convex case]\label{th:rand_VC}
Suppose that $d_{VC} < +\infty$. If
\begin{equation} \label{eq:findN2}
N \geq \frac{5 \eps}{(\eps-\alpha)^2} \left[ d_{VC} \ln \left( \frac{40 \eps }{(\eps-\alpha)^2} \right) + \ln \frac{4}{\beta} \right],
\end{equation}
then the randomized solution $(\vartheta^\ast, \lambda^\ast)$ is such that
\begin{align*}
\Prob \left\{ D \left( y^\sys,y^\m_{\lambda^\ast} \right)^2 \leq h_{\vartheta^\ast}(x_0) \right\} \geq
1 - \eps
\end{align*}
with confidence at least $1-\beta$. \hfill \qed
\end{theorem}
\textbf{Proof:} see \ref{appendix:2}. \\
\\
The interpretation of Theorem~\ref{th:rand_VC} is the same as for Theorem~\ref{th:scenario}, and, likewise, all the comments we made before still apply. \\

Note that although the lack of convexity makes the resolution of optimization problems in Algorithm~\ref{algo:scenario_algo} harder,
the guarantees provided by Theorem~\ref{th:rand_VC} apply to any local solution, so that one has not necessarily to head for the global optimizer when solving the optimization problems in Algorithm~\ref{algo:scenario_algo}.
In turn, though the assumption that the VC dimension is finite is relatively mild, the computation of $d_{VC}$ is nontrivial and often only conservative bounds can be derived. This means that the sample size $N$ in Theorem~\ref{th:rand_VC} is overestimated, with an increase of the computational complexity that can hamper the applicability of the approach. In this respect, the computation of tight bounds for $d_{VC}$ is still an open issue. Results for specific classes of problems are available in \cite{ATC2009} and references therein.

\section{Jump Linear Stochastic Systems: A numerical example}\label{sec:example}

\begin{figure}[H]
\begin{centering}
\includegraphics[width=0.5\paperwidth]{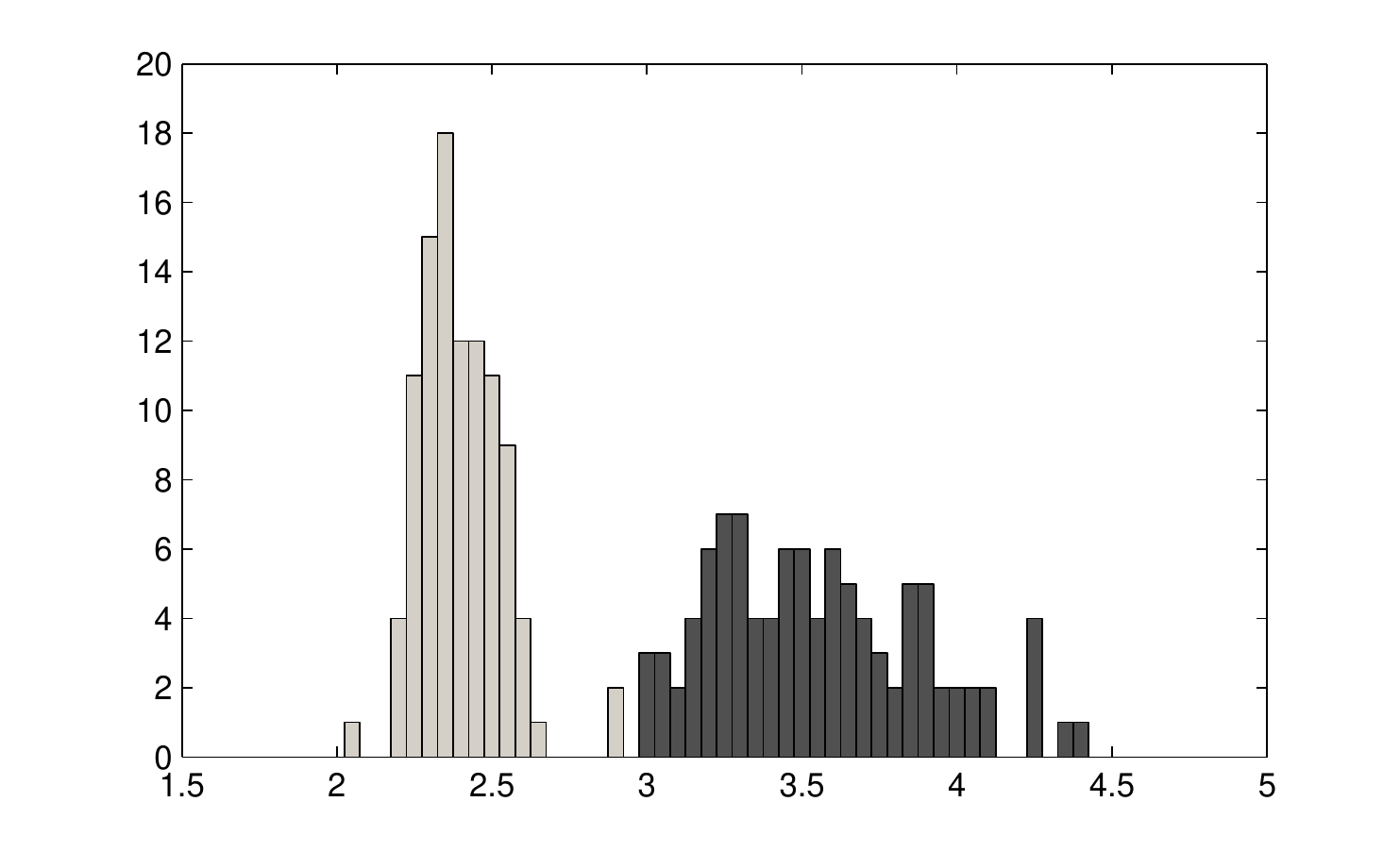}\caption{Histograms showing the performance of the randomized solution to \eqref{eq:cc-opt} when $\alpha=0.10$.\label{fig:histo0.1}}
\par\end{centering}
\end{figure}

In this section, we illustrate the results obtained by the proposed randomized method on a numerical example that was first studied in \cite{JP09}, and we compare them with those obtained by the stochastic bi-simulation method revised at the end of Section \ref{sec:problem}. \\
Suppose that system $S$ is a JLSS whose state $x_t \in
\mathbb{R}^6$ is governed by the SDE in equation \eqref{eq:jump-diffusion} with
\begin{align*}
& A = \begin{bmatrix}
-1 & -10 & 0 & 0 & 0 & 0 \\
10 & -1 & 0 & 0 & 0 & 0 \\
0 & 0 & -2 & -20 & 0 & 0 \\
0 & 0 & 20 & -1 & 0 & 0 \\
0 & 0 & 0 & 0 & -2 & 0\\
0 & 0 & 0 & 0 & 0 & -2.5
\end{bmatrix}, &
F = 0.5 \cdot \begin{bmatrix}
1 & 0 & 0 & 0 & 1 & 1\\
0 & 1 & 0 & 0 & 0 & 0\\
0 & 0 & 1 & 0 & 1 & 1\\
0 & 0 & 0 & 1 & 0 & 0\\
1 & 0 & 0 & 0 & 0 & 1\\
0 & 0 & 1 & 0 & 1 & 0
\end{bmatrix},
\end{align*}
in-between the jump times of a Poisson process with rate $\nu=0.5$.
At the jump times the state is reset according to \eqref{eq:reset} where $R=0.7\cdot I_{6}$, $I_m$ denoting the identity matrix of size $m$.
The output of interest $y^\sys_t$ takes values in $\mathbb{R}^2$ and is given by $y^\sys_t = C x^\sys_t$, where
\begin{align*}
C=\begin{bmatrix}0.84 & -1.03 & 1.07 & -0.88 & 0.5 & 0\\
-0.6 & -1.35 & -0.26 & -0.27 & 0 & -0.5
\end{bmatrix}.
\end{align*}
To the purpose of reproducing the output $y^\sys_t$ along the time horizon $[0,10]$, we consider three different JLSS models as indicated in Example \ref{example:JLSS}, equations \eqref{eq:jump-diffusion-bis}-\eqref{eq:output-bis}:
\begin{itemize}
\item
model $M_1$ is obtained by considering only the first four state variables in $x^\sys_t$ and re-defining the matrices entering the JLSS definition of system $S$ by removing those rows/columns that relates to the contribution of the last two state variable in $x^\sys_t$. To be precise,
\begin{align*}
& \tilde A = \begin{bmatrix}
-1 & -10 & 0 & 0 \\
10 & -1 & 0 & 0  \\
0 & 0 & -2 & -20 \\
0 & 0 & 20 & -1
\end{bmatrix}, & \tilde F = 0.5 \cdot I_4 \\
& \tilde C = \begin{bmatrix}0.84 & -1.03 & 1.07 & -0.88\\
-0.6 & -1.35 & -0.26 & -0.27
\end{bmatrix} & \tilde R = 0.7 \cdot I_4.
\end{align*}

\item model $M_2$ is obtained by removing the contribution of the Brownian motion, i.e. by letting $\tilde A = A$, $\tilde F = 0$, $\tilde R = R$, $\tilde C = C$.

\item model $M_3$ is obtained by removing the contribution of the Poisson process, i.e. by letting $\tilde A = A$, $\tilde F = F$, $\tilde R = 0$, $\tilde C = C$.
\end{itemize}
As for models $M_2$ and $M_3$, the initial state $x_0$ of system $S$ is mapped into that of the approximating models through the identity map, whereas the initial state of model $M_1$ is given by the first four entries of vector $x_0$. The performance of each model $M_i$, $i=1,2,3$, as an abstraction of $S$ is assessed through the following chance-constrained optimization problem
\begin{align}
&\min_{\Theta \succeq 0}\,
\trace \left(\Theta \E
\begin{bmatrix}x_0 {x_0}' &{x_0} \\{x_0}' & 1
\end{bmatrix} \right) \label{eq:cc_example} \\
&\text{\rm subject to: }
\Prob\left\{ \Big(\sup_{t\in[0,T]}||y^\sys_t-y^{\m_i}_t||\Big)^{2}
\leq
\begin{bmatrix}{x_0}' & 1\end{bmatrix}
\Theta
\begin{bmatrix}x_0\\
1
\end{bmatrix}\right\} \geq1-\epsilon. \nonumber
\end{align}
Problem \eqref{eq:cc_example} was approximately solved by means of Algorithm~\ref{algo:scenario_algo} and its further variants with the random and block constraint removal rules implemented (see Remark \ref{rmk:alternative_algo}). We set $\eps=0.25$, $\beta=10^{-10}$, and progressively increase $\alpha$ from $0.10$ to $0.22$. Correspondingly, according to Theorem~\ref{th:scenario}, $N$ grows from $1697$ to $91786$  (note that $r=28$ since $\Theta$ is a $7 \times 7$ symmetric matrix). We adopted the greedy removal for $\alpha=0.10$, random removal for $\alpha=0.15$ and block removal for larger values of $\alpha$.
For the sake of comparison the stochastic bi-simulation function method was also used. Results obtained when the state $x_0$ is Gaussian with zero mean and identity covariance ($x_0\sim\mathcal{N}(0,I_6)$) are shown in Table \ref{table:normal_assessment}.
\begin{table}[H]
\begin{centering}
\begin{tabular}{|>{\raggedright}m{2.5cm}|c|c|c|c|c|c|}
\hline
$x_{0}\sim\mathcal{N}(0,I_6)$ & \multirow{1}{*}{$\alpha=0.10$} & \multirow{1}{*}{$\alpha=0.15$} & \multirow{1}{*}{$\alpha=0.20$}& \multirow{1}{*}{$\alpha=0.22$} & \multirow{1}{*}{SSF}\tabularnewline
\hline
\multirow{2}{2.4cm}{\quad \quad $M_1$} & \multirow{1}{*}{$J=4.86$} & $J=3.515$ & $J=2.88$ & $J=2.59$ & $J=10.13$\tabularnewline
\cline{2-6}
 & $\hat\eps=0.127$ & $\hat\eps=0.160$ & $\hat\eps=0.200$& $\hat\eps=0.222$& $\hat\eps=0.040$\tabularnewline
\hline
\multirow{2}{2.4cm}{\quad\quad $M_2$} & $J=15.26$ & $J=10.97$& $J=8.42$& $J=7.45$ & $J=19.77$\tabularnewline
\cline{2-6}
 & $\hat\eps=0.121$  & $\hat\eps=0.165$ & $\hat\eps=0.206$ & $\hat\eps=0.222$ & $\hat\eps=0.107$\tabularnewline
\hline
\multirow{2}{2.4cm}{\quad\quad $M_3$} & $J=16.99$ & $J=10.43$ & $J=7.22$ & $J=6.20$ & $J=15.63$\tabularnewline
\cline{2-6}
 &$\hat\eps=0.118$ & $\hat\eps=0.158$ & $\hat\eps=0.203$ & $\hat\eps=0.224$& $\hat\eps=0.132$\tabularnewline
\hline
\end{tabular}
\par\end{centering}
\caption{Performance of the randomized method and of the stochastic bi-simulation function (SSF) method, when  $x_0$ is a Gaussian random variable with zero mean and identity covariance.\label{table:normal_assessment}}
\end{table}
In this table, $J$ denotes $\E [h_{\vartheta^\ast}(x_0)]$, i.e. the average upper bound on $(\sup_{t\in[0,T]}||y^\sys_t-y^{\m_i}_t||)^{2}$ in correspondence of the found solution (see \eqref{eq:cc_example} and  \eqref{eq:SSF-accuracy} for the expression of $h_\vartheta(x_0)$ in the randomized approach and in the bi-simulation function method). Instead, $\hat \eps$ is a Monte Carlo estimate of the actual violation probability. As expected $\hat \eps$ is below the threshold $\eps=0.25$ in all cases. \\
The table shows that the average accuracy $J$ provided by the stochastic bi-simulation function method is typically worse than that obtained by the randomized method. Consistently with this result, in the stochastic bi-simulation function method $\hat\eps$ is significantly lower than the desired value $\eps$, especially in the case of model $M_1$. \\
As for the randomized method, irrespectively of the greedy, random or block implementation, $\hat\eps$ is close to the empirical violation $\alpha$. If $\alpha$ is increased, then $J$ improves and $\hat\eps$ grows. This is a strength of the proposed approach, where, by means of the choice of $\alpha$, the user can modulate the actual violation probability so as to better match the desired $\eps$ value. The stochastic bi-simulation function method, instead, does not offer this opportunity and generally provides conservative values for the average accuracy $J$. \\
In order to assess the conservativeness of the proposed randomized approach for the chosen parametrization $h_\vartheta(x_0)$ of the accuracy function, we considered model $M_1$ and the solution $\theta^\ast$ obtained for $\alpha=0.22$ and determined the empirical density with respect to $1000$ extractions of $x_0$ of
\begin{equation} \label{eq:max_dev}
\max \left\{ \left[ h_{\vartheta^\ast}(x_0)-(\sup_{t\in[0,T]}||y^\sys_t-y^{\m_i}_t||)^{2} \right]^+ \right\},  \quad ([\cdot]^+ = \mbox{positive part})
\end{equation}
where $\max$ was in turn empirically determined over $4500$ realization of $w$.
\begin{figure}[H]
\begin{centering}
\hspace*{-7mm}\includegraphics[width=0.35\paperwidth]{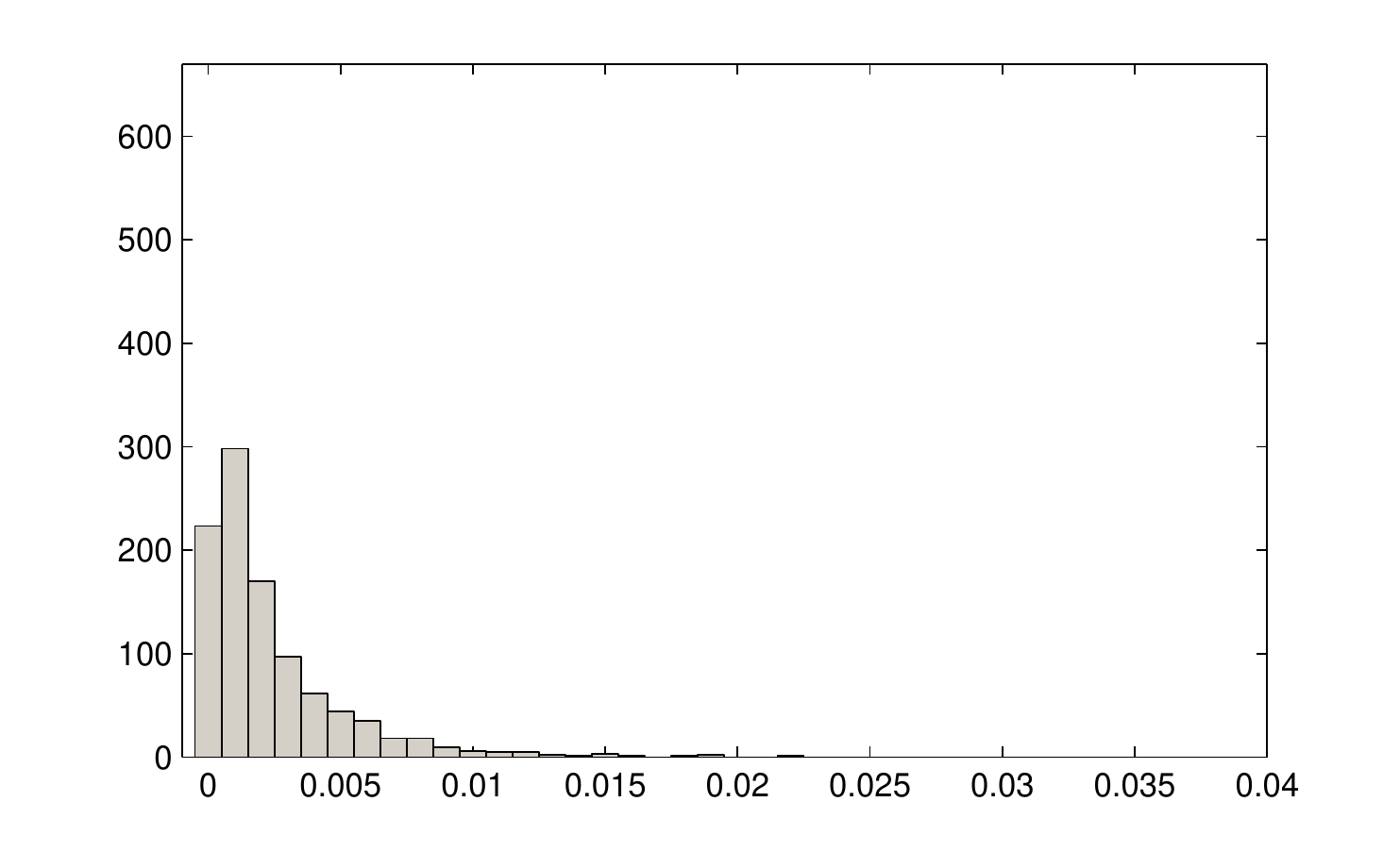}\includegraphics[width=0.35\paperwidth]{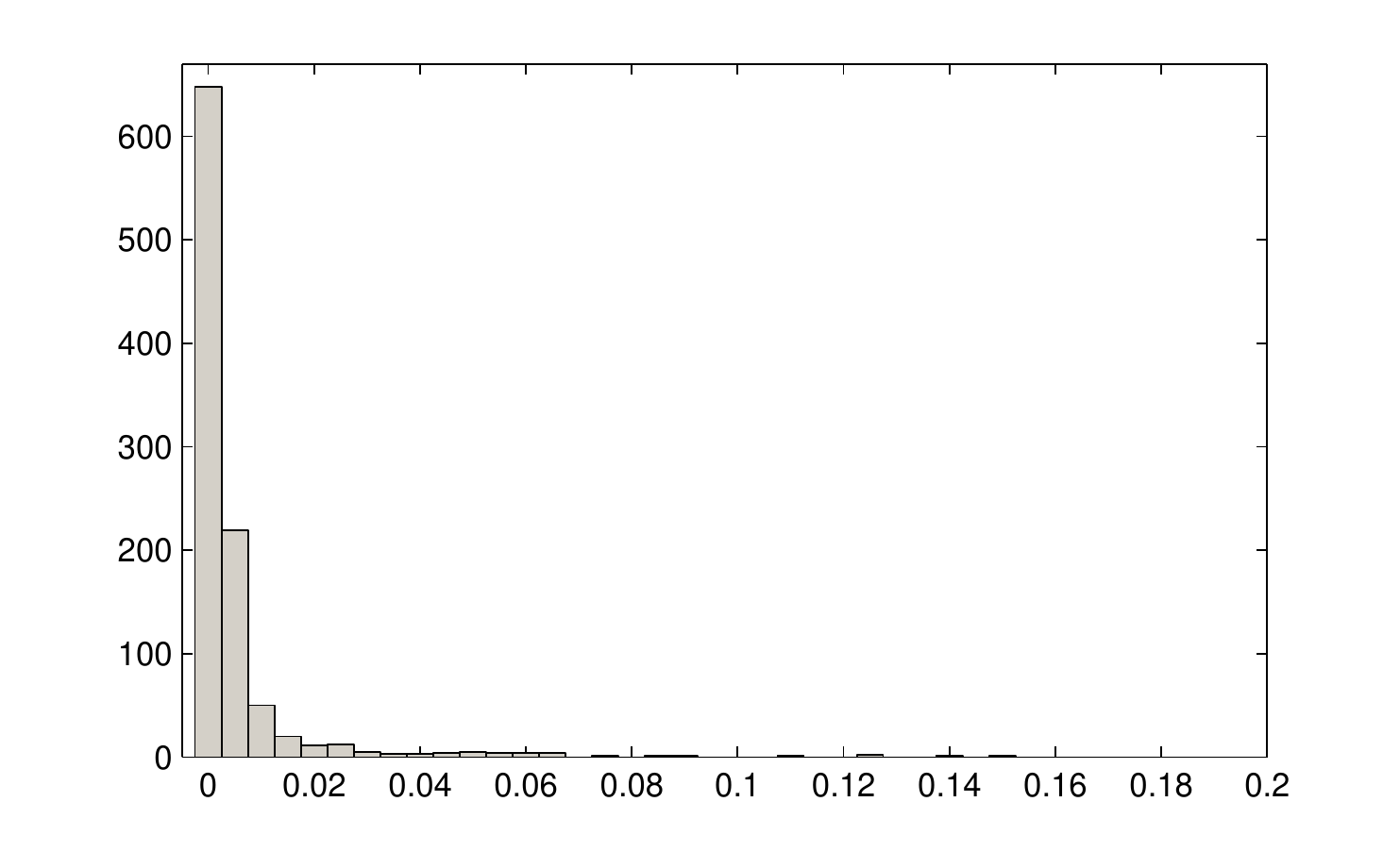}
\caption{Histograms showing the performance of the randomized solution to \eqref{eq:cc_example} for model $M_1$ when $\alpha=0.22$.\label{fig:assessment}}
\par\end{centering}
\end{figure}
The corresponding histogram is depicted in Figure \ref{fig:assessment}, together with its normalized version where \eqref{eq:max_dev} is divide by the value of $(\sup_{t\in[0,T]}||y^\sys_t-y^{\m_i}_t||)^{2}$ corresponding to the maximum.
\\
\\
Suppose now that we want to optimize the initialization for the model $M$ so as to better reproduce the system output.
More precisely, we want to optimize the $\tilde n \times n$ matrix $L$ mapping the initial state $x_0$ of $S$ into the initial state $x^\m_0$ of $M$: $x^\m_0 = L x_0$. \\
Given that the JLSS is characterized by linear drift and diffusion terms and by a linear reset map, it is easily seen that the resulting function $\fmod_\lambda(x_0,w)$ is linear in $\lambda =L$, so that $D\left(y^\sys,y^\m \right)^2$ is convex in $\lambda$, and Theorem ~\ref{th:scenario} can be applied to optimize the performance of $M$ with respect to $\lambda$ (and $\vartheta$). \\
Before providing details on how to implement the randomized solution to the
chance-constrained optimization, we need to specify how to
determine $y^\m = \fmod_\lambda(x_0,w)$ as an explicit function
of $\lambda = L$ for each pair of initial condition $x_0$ and
input realization $w$. To this purpose one can simulate $\tilde
n$ executions of equations \eqref{eq:jump-diffusion-bis} and
\eqref{eq:reset-bis}, each with the same input $w$ and for the
$\tilde n$ initial conditions $x^\m_0 = e_1, \dots, x^\m_0 =
e_{\tilde n}$, where $e_i$ is the vector with all elements
equal to 0 except for the $i$-th element equal to 1. Then,
$y^\m$ can be obtained as a linear combination of these
executions according to $L x_0$. More precisely, letting
$\xi_{i,t}$ be the execution of \eqref{eq:jump-diffusion-bis}
and \eqref{eq:reset-bis} associated with the initial condition
$e_i$ at time $t$, and letting
\begin{align*}
\Xi_t = [\xi_{1,t} \; \xi_{2,t} \; \cdots \xi_{{\tilde n},t}]
\end{align*}
be the matrix with $\xi_{i,t}$ as columns, then we have that $y^\m_t = \tilde C \Xi_t L x_0$, $\forall t\in [0,T]$. \\
This eventually leads to the following steps:

\begin{itemize}
\item[1:] Choose $\eps \in (0,1)$, $\beta \in (0,1)$, and $\alpha \in
[0,\eps)$. Let $N$ be the smallest integer satisfying \eqref{eq:findN_implicit} (or, alternatively, let $N$ be the smallest integer satisfying \eqref{eq:findN}).

\item[2:] Extract $N$ realizations of the stochastic input $w^{(i)}$,
$i=1,2,\dots,N$, and of the initial condition $x_0^{(i)}$,
$i=1,2,\dots,N$.

\item[3:] Run the corresponding $N$ executions of $S$ to compute
via simulation $N$ realizations of the system output
\begin{align*}
y^{\sys,(i)}=\fsys(x_0^{(i)},w^{(i)}),  i=1,2, \dots,N.
\end{align*}

\item[4:] \emph{for} $i=1,\ldots,N$
\begin{itemize}
\item[] Run $\tilde n$ executions of
    \eqref{eq:jump-diffusion-bis} and \eqref{eq:reset-bis}
    with varied initial conditions $x^\m_0 =
    e_1,\ldots,x^\m_0=e_{\tilde n}$ and same input equal to
    $w^{(i)}$ so as to generate $\Xi^{(i)}_t$.
\end{itemize}
\emph{end}

\item[5:] Run Algorithm~\ref{algo:scenario_algo} with the following objective
\begin{align*}
\min_{\Theta \succeq 0, L} \trace \left(
\Theta \E\begin{bmatrix}x_0 {x_0}' &{x_0} \\
{x_0}'& 1 \end{bmatrix} \right)
\end{align*}
and constraints
\begin{align*}
& D(y^{\sys,(i)},y^{\m,(i)}_L)^2 \leq {\begin{bmatrix} {x_0^{(i)}}' & 1\end{bmatrix}}
\Theta
\begin{bmatrix}x_0^{(i)} \\ 1 \end{bmatrix}
\end{align*}
where $y^{\m,(i)}_{L,t} = \tilde C \Xi^{(i)}_t L x_0^{(i)}, \quad t
\in [0,T]$.
\end{itemize}
We next report the results obtained when the initial state of $S$ is deterministic and given by $x_0={[1\, 1 \,1 \,1 \,1 \,1]}'$, and $T=0.2$. In this case, the accuracy function can be replaced by a scalar $h$. \\
The randomized method with random constraint removal was run with the following set of parameters:  $\eps=0.25$, $\beta=10^{-10}$ and $\alpha=0.10$.
As in the performance assessment case, the obtained solution $(h^\star,L^\star)$ is such that the actual violation probability $\hat\eps$ is close to the empirical violation $\alpha$.
\\
Figure \ref{fig:histo0.1} represents two histograms: the gray histogram refers to the values of  $h$ obtained by computing 100 times the randomized solution to \eqref{eq:cc-opt} when both $x^{\m_1}_0=Lx_0$ and $h$ are optimized, whereas the black histogram refers to the case when only $h$ is optimized and
$x_0^{\m_1}$ is set equal to the first four components of $x_0$: $x_0^{\m_1}={[1\, 1 \,1 \,1]}'$. The optimization of the initial condition is shown to be quite effective in improving the accuracy of the abstracted model, despite of the randomness affecting the solution. \\
In order to improve the solution one should adopt a larger value for $\alpha$, say $\alpha=0.22$, thus getting the actual violation probability close to the desired $\eps=0.25$ value. This may, however, cause an excessive computational effort.
To cope with this issue, one can adopt a two-step procedure similar to that discussed in \cite{Care2011,Care2014}, where first $\alpha=0.10$  is used to optimize both $x_0^{\m_1}$ and $h$, and then for $\alpha=0.22$ only $h$ is optimized setting $x_0^{\m_1}$ equal to the value $x_0^{\m_1,\star}$ obtained when $\alpha=0.10$. The guarantees provided by Theorem \ref{th:scenario} on the re-optimized $h$  still hold for model $M_1$ initialized with $x_0^{\m_1,\star}$. The value for $h$ obtained through this 2-step procedure is better than that obtained by setting $x_0^{\m_1}$ equal to the first 4 components of $x_0$, i.e., $x_0^{\m_1}={[1\, 1 \,1 \,1]}'$, and optimizing $h$ with $\alpha=0.22$. This is shown  in Figure \ref{fig:histo0.22}, where the histograms of $h$ obtained by running 100 times the 2-step procedure (gray histogram) and by optimizing only $h$ with $x_0^{\m_1}={[1\, 1 \,1 \,1]}'$ (black histogram) are depicted. This shows that the optimization of $x_0^{\m_1}$ leads to an improved accuracy $h$, even when performed according to the suggested 2-step heuristics.
\begin{figure}[H]
\begin{centering}
\includegraphics[width=0.5\paperwidth]{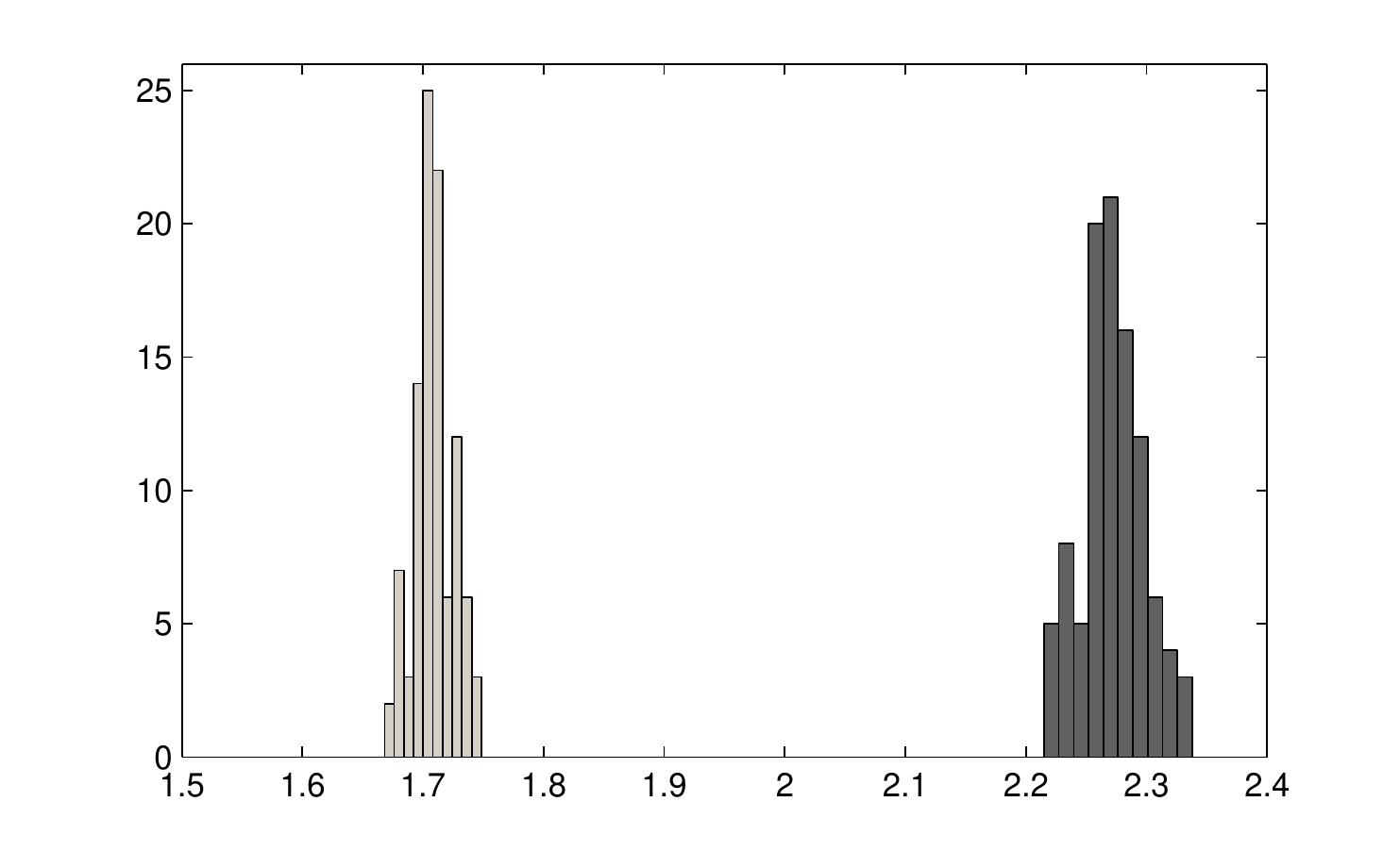}\caption{Histograms showing the performance of the randomized solution to \eqref{eq:cc-opt} when $\alpha=0.22$.\label{fig:histo0.22}}
\par\end{centering}
\end{figure}
Note that the histograms in Figure \ref{fig:histo0.1} reveal a larger variability compared to those in Figure \ref{fig:histo0.22}. This is due to the fact that the latter results are associated to a value of $\alpha$ that is closer to the desired $\eps$ value for the violation probability.

\section{Conclusions}\label{sec:conclusions}

In this paper, we proposed a simulation-based method for the analysis and design of an approximate abstraction of a SHS. This approach rests on recent results on the randomized solution to chance-constrained programs, and turns out to be much less conservative than other approaches in the literature. The counterpart for the improved performance is that guarantees on the quality of the solution hold with a certain confidence,
which, however, can be set arbitrarily close to 1, though at the expense of a larger computational effort.  \\
A key advantage of the proposed method is that it does not require specific assumptions on the system $S$ to be approximated. In the case of performance assessment, a computational convenient convex formulation is also suggested.
Since some of the approaches in the literature to the design of simpler abstracted models of a hybrid system do not provide an evaluation of the model accuracy, see e.g. \cite{MSVBB08}, the proposed reformulation can then be used to complement them with such an evaluation, \cite{PP2014}. \\
Our method can also be employed in principle  to design optimal abstracted models, in that it allows to choose the best model in some given parameterized model class. The quite challenging issue of choosing the best model class, however, remains open.

\appendix

\section{Proof of Corollary \ref{corol:scenario}} \label{appendix:1}

Corollary \ref{corol:scenario} is proven by following chain of implications, which shows that \eqref{eq:findN} implies \eqref{eq:findN_implicit} in Theorem \ref{th:scenario}.
$$
\begin{array}{c}
N \geq \frac{(2+\alpha) \eps}{(\eps-\alpha)^2} \left[ (r-1) \ln \left( \frac{2 \eps (2+\alpha)  (r-1)}{(\eps-\alpha)^2} \right) + \ln \frac{1}{\beta} \right] + \frac{r-1}{2}\\
\Downarrow \\
N \geq \frac{2 \eps}{(\eps-\alpha)^2} \ln \frac{1}{\beta} + \frac{2 \eps (r-1)}{(\eps-\alpha)^2} \ln \left( \frac{2\eps (2+ \alpha)  (r-1)}{(\eps-\alpha)^2} \right) + \frac{r-1}{2+\alpha} + \frac{\alpha}{2+\alpha} N \\
\Downarrow \\
N \geq  \frac{2(r-1)}{\eps-\alpha} + \frac{2 \eps}{(\eps-\alpha)^2} \ln \frac{1}{\beta} + \frac{2 \eps (r-1)}{(\eps-\alpha)^2} \ln \left( \frac{2 \eps(2+ \alpha) (r-1)}{(\eps-\alpha)^2} \right) + \frac{1}{2+\alpha} \left( \alpha N + r-1 - \frac{2\eps (2+ \alpha) (r-1)}{(\eps-\alpha)^2} \right)\\
\Downarrow \\
( \mbox{since } \ln(x) \leq \ln(\bar x) +\frac{1}{\bar x} (x - \bar x) \mbox{ for any } \bar x )\\
\Downarrow \\
N \geq  \frac{2(r-1)}{\eps-\alpha} + \frac{2 \eps}{(\eps-\alpha)^2} \ln \frac{1}{\beta} + \frac{2 \eps (r-1)}{(\eps-\alpha)^2} \ln \left( \alpha N + r-1 \right)\\
\Downarrow \\
\frac{(\eps-\alpha)^2 N}{2\eps} - \frac{(\eps-\alpha) (r-1)}{\eps} \geq  \ln \frac{\left( \alpha N + r-1 \right)^{r-1}}{\beta} \\
\Downarrow \\
\frac{\left( (\eps-\alpha)N-r+1 \right)^2}{2\eps N} \geq  \ln \frac{\left( \alpha N + r -1 \right)^{r-1}}{\beta} \\
\Downarrow \\
\left( \alpha N + r -1\right)^{r-1} \mathrm{e}^{-\frac{\left( (\eps-\alpha)N-r+1 \right)^2}{2\eps N}} \leq  \beta \\
\Downarrow \\
\left( \lfloor \alpha N \rfloor + r -1\right)^{r-1} \mathrm{e}^{-\frac{\left( \eps N - \left( \lfloor \alpha N \rfloor + r -1\right) \right)^2}{2\eps N}} \leq  \beta \\
\Downarrow \\
{\lfloor \alpha N \rfloor + r -1\choose \lfloor \alpha N \rfloor} \mathrm{e}^{-\frac{\left( \eps N - \left( \lfloor \alpha N \rfloor + r -1\right) \right)^2}{2\eps N}} \leq  \beta \\
\Downarrow \\
(\mbox{by using the Chernoff inequality, \cite{TCD2013}}) \\
\Downarrow \\
{\lfloor \alpha N \rfloor + r -1\choose \lfloor \alpha N \rfloor} \sum_{i=0}^{\lfloor \alpha N \rfloor + r-1} {N \choose i} \eps^i (1-\eps)^{N-i} \leq \beta
\end{array}
$$

\section{Proof of Theorem \ref{th:rand_VC}} \label{appendix:2}

By Theorem 7 in \cite{ATC2009}, condition \eqref{eq:findN2} implies that
\begin{equation} \label{eq:B1}
\frac{1}{N} \sum_{i=1}^N \mathbf{1}_{D \left( y^{\sys,(i)},y^{\m,(i)}_{\lambda}\right)^2 > h_{\vartheta}(x_0^{(i)})} \leq \alpha
\end{equation}
and
\begin{equation} \label{eq:B2}
\Prob \left\{ D \left( y^\sys,y^\m_{\lambda^\ast} \right)^2 \leq h_{\vartheta^\ast}(x_0) \right\} > \frac{1}{N} \sum_{i=1}^N \mathbf{1}_{D \left( y^{\sys,(i)},y^{\m,(i)}_{\lambda}\right)^2 > h_{\vartheta}(x_0^{(i)})} + \eps - \alpha \\
\end{equation}
hold simultaneously with probability smaller than $\beta$. \\
Since, by construction,
$$
\frac{1}{N} \sum_{i=1}^N \mathbf{1}_{D \left( y^{\sys,(i)},y^{\m,(i)}_{\lambda} \right)^2 > h_{\vartheta}(x_0^{(i)}) } = \frac{\lfloor \alpha N \rfloor}{N} \leq \alpha,
$$
\eqref{eq:B1} and \eqref{eq:B2} imply that
$$
\Prob \left\{ D \left( y^\sys,y^\m_{\lambda^\ast} \right)^2 > h_{\vartheta^\ast}(x_0) \right\} > \eps
$$
with probability smaller than $\beta$, which is equivalent to say that
$$
\Prob \left\{ D \left( y^\sys,y^\m_{\lambda^\ast} \right)^2 \leq h_{\vartheta^\ast}(x_0) \right\} \geq 1 - \eps
$$
with confidence at least $1-\beta$.

\end{document}